\documentclass[conference,a4paper]{IEEEtran}

\usepackage{algorithm,algorithmic}
\usepackage{amsmath,amssymb,mathrsfs,amsbsy}
\usepackage{cite}
\usepackage{subfigure}
\usepackage{graphicx,color}

\usepackage{hyperref}
\usepackage{subfigure}
\usepackage{balance}
\usepackage{tikz,siunitx}
\usepackage{url}
\usepackage[nomain,acronym,toc]{glossaries}

\newtheorem{defn1}{Definition}
\newtheorem{lem}{Lemma}

\newtheorem{thm}{Theorem}

\newacronym{urllc}{URLLC}{ultra-reliable low-latency communication}
\newacronym{itu}{ITU}{international telecommunication union}
\newacronym{ccdf}{CCDF}{complementary cumulative distribution function}
\newacronym{cdf}{CDF}{cumulative distribution function}
\newacronym{snr}{SNR}{signal-to-noise ratio}
\newacronym{ofdma}{OFDMA}{orthogonal frequency-division multiple-access}
\newacronym{se}{SE}{spectral efficiency}
\newacronym{fdma}{FDMA}{frequency-division multiple-access}
\newacronym{csit}{CSIT}{channel-state-information at the transmitter}
\newacronym{mac}{MAC}{medium access control}
\newacronym{tti}{TTI}{transmission time interval}
\newacronym{harq}{HARQ}{hybrid automatic repeat request}
\newacronym{isnr}{iSNR}{instantaneous signal-to-noise-ratio}

\definecolor{sblue}{RGB}{0,51,120}
\definecolor{sred}{RGB}{200,51,130}

\ifCLASSINFOpdf
\else
\fi

\begin{document}

\title{Power Allocation for FDMA-URLLC Downlink with Random Channel Assignment}

\author{Jinfei Wang, Yi Ma, Na Yi, and Rahim Tafazolli\\
	{\small 5GIC and 6GIC, Institute for Communication Systems, University of Surrey, Guildford, UK, GU2 7XH}\\
	{\small Emails: (jinfei.wang, y.ma, n.yi, r.tafazolli)@surrey.ac.uk}
}
\markboth{}%
{}

\maketitle

\begin{abstract}
Concerning ultra-reliable low-latency communication (URLLC) for the downlink operating in the frequency-division multiple-access with random channel assignment, a lightweight power allocation approach is proposed to maximize the number of URLLC users subject to transmit-power and individual user-reliability constraints.	
Provided perfect channel-state-information at the transmitter (CSIT), the proposed approach is proven to ensure maximized URLLC users. Assuming imperfect CSIT, the proposed approach still aims to maximize the URLLC users without compromising the individual user reliability by using a pessimistic evaluation of the channel gain.
It is demonstrated, through numerical results, that the proposed approach can significantly improve the user capacity and the transmit-power efficiency in Rayleigh fading channels. With imperfect CSIT, the proposed approach can still provide remarkable user capacity at limited cost of transmit-power efficiency.
\end{abstract}

\section{Introduction}\label{secI}
The power-allocation problem in ultra-reliable low-latency communication (URLLC) could need fundamental redesign. 
To meet the stringent latency constraint (e.g., $0.5$ ms of transmission time or less \cite{8472907}), the access point (AP) needs to deliver each packet within limited shots of transmissions \cite{9114878}. This indicates the packet error rate (PER) cannot be averaged on the time domain. Moreover, the PER cannot be averaged over users, since URLLC is interested in each individual user's reliability. 
Instead, the AP should be aware of the PER for each packet based on the evaluation of instant signal-to-noise ratio (iSNR) \cite{8660712,9120745,wang2021massivemimo}.
This makes power-allocation design for URLLC fundamentally different from throughput-oriented systems that usually aim at sum-rate maximization \cite{9044874,1427697}, mean-square-error minimization \cite{1687745}, min-capacity maximization \cite{9049046}, capacity-constrained energy-efficiency \cite{8693052} and so on.
On the other hand, the power consumption to achieve ultra-reliability can be prohibitively high when the wireless channel is deeply faded. 
Current works for URLLC focus on utilizing diversity gain to lower the probability of deep fading (e.g., \cite{8660712,9120745,wang2021massivemimo}).
However, this is not suitable for narrow-band users, who can hardly take advantage of channel spatial and frequency diversities.
In this case, the power should be dynamically allocated to compensate the loss in iSNR and enable more users to enjoy URLLC service. All these discussions motivate us to look into the power-allocation problem of URLLC.

In this paper, we are interested in a special URLLC power allocation problem, where the transmitter communicates to a set of narrow-band receivers in the frequency-division multiple-access (FDMA) downlink \footnote{FDMA is considered for concurrent and orthogonal multiuser transmissions; and this model is well suitable for OFDMA systems.}\cite{9129054}. We assume single-antenna transmission and reception so as to get the research problem more focused (e.g., \cite{9378575,9416453}).
Moreover, it is assumed that the narrow-band sub-channels are randomly assigned to the receivers. This is because the random channel-assignment protocol is more latency friendly than other sophisticated medium access control (MAC) protocols. This is important for URLLC apart from low physical-layer latency \cite{8636206}. 
In such a system, some sub-channels may not be able to provide URLLC service due to deep fading, while the others only need part of their power to provide URLLC service. This calls for a dynamic power allocation approach to enable more sub-channels to deliver URLLC service. For the sake of simplicity, we define the users whose sub-channels can provide URLLC service as URLLC users. Therefore, the aim of power allocation is to maximize the number of URLLC users in the system (i.e., user capacity) subject to: 1) the overall power constraint; and 2) the individual user reliability constraint.
Our analysis proves that when provided perfect channel-state-information at the transmitter (CSIT), this problem can be solved by a lightweight power-sorting approach proposed by us.
When the CSIT is imperfect, the transmitter faces additional uncertainty (i.e., outage probability) and each sub-channel needs more power to provide URLLC service. 
By increasing the transmission power based on a pessimistic evaluation of the channel gain, it is shown that the proposed lightweight sorting approach still aims to maximize the URLLC users without compromising the individual user reliability. 

Numerical results reveal that the proposed power-sorting approach can offer significant improvements on user capacity as well as transmit-power efficiency in Rayleigh fading channels.
Provided perfect CSIT, the power-sorting approach can provide an improvement on user capacity of at least $0.12$ (out of $1$) compared to equal-power allocation, waterfilling allocation or equal-iSNR allocation.
There is also more than $10$-dB improvement on average URLLC-user power consumption for the proposed approach.
Moreover, both improvements are more significant when the average power per sub-channel becomes smaller.
When the channel knowledge error is white Gaussian \cite{1427697}, the power-sorting approach can preserve remarkable performance (less than a degradation of $0.07$) with an average power-consumption increase of less than $2.5$ dB.

\section{System Model and Problem Formulation}
Consider a downlink FDMA-URLLC system with $M$ narrow-band sub-channels, where the channel is assumed to be static in one transmission. Denote the channel coefficient and channel gain of the $m^\mathrm{th}$ sub-channel ($m=0,1,...,M-1$) by $h_m$ and $a_m$ ($a_m\triangleq|h_m|^2$), respectively. Each sub-channel is assigned randomly to a user that may require URLLC service. Hence, each sub-channel is assumed to be independently fading with respect to other sub-channels. The average available power per sub-channel is denoted by $P$, so the total available power of the system is $MP$.

When the $m^\mathrm{th}$ sub-channel is enabled to provide URLLC service, a short-length packet ($\mathbf{x}_m$) of $L$ symbols is transmitted to the assigned user with a PER requirement $\overline{\mathscr{P}}_\mathrm{err}$. Then, the received signal $\mathbf{y}_m$ is given by
\begin{equation}\label{eq01}
\mathbf{y}=h_m\sqrt{P_m}\mathbf{x}_m+\mathbf{v}_m,
\end{equation}
where $P_m$ stands for the allocated power to the $m^\mathrm{th}$ sub-channel and $\mathbf{v}_m$ for additive white Gaussian noise ($\mathbb{E}(\mathbf{v}\mathbf{v}^H=N_0\mathbf{I})$). Here $(\cdot)^H$ stands for the Hermitian of a vector. 
Different from conventional throughput-oriented systems that use Shannon capacity for analysis, the packet $\mathbf{x}_m$ is in the finite-length regime, and the decoding error rate ($\epsilon$) is therefore not negligible \cite{8472907}. 
In this case, the Polyanskiy's finite-length achievable rate is used for our study \cite{224132864}:
\begin{equation}\label{eq02}
R_{m}(P_m)=\log_2\left( 1+\frac{a_mP_{m}}{N_0} \right)-\sqrt{\frac{V_{m}}{L}}\frac{\mathcal{Q}^{-1}(\epsilon)}{\ln 2},
\end{equation}
where $\mathcal{Q}$ stands for the Gaussian tail function and $V_m$ for the channel dispersion given by
\begin{equation}\label{eq03}
V_m=1-\left(1+\frac{a_mP_m}{N_0}\right)^{-2}.
\end{equation}

To fulfill the URLLC requirement, $R_m$ needs to reach the required transmission rate $\overline{R}$ (determined by the length of $\mathbf{x}_m$ and latency constraint), such that the decoding error rate $\epsilon$ would not exceed $\overline{\mathscr{P}}_\mathrm{err}$.
Moreover, the error of transmission can also occur due to channel outage. This happens particularly when the AP is not perfectly aware of the channel (i.e., $h_m$ and $a_m$).
Specifically for URLLC, the definition of outage probability ($\mathscr{P}_\mathrm{out}$) is based on the finite-length achievable rate rather than Shannon capacity:
\begin{equation}\label{eq04}
\mathscr{P}_\mathrm{out}\triangleq\mathscr{P}(R_m(P_m)<\overline{R}).
\end{equation}
In general, it is assumed that the packet is successfully delivered only when the channel is not in outage, and $\mathbf{x}_m$ is successfully decoded as well (e.g., \cite{9120745,wang2021massivemimo,Wang2022}). Hence, when ultra-reliability is achieved, the following inequality holds: 
\begin{IEEEeqnarray}{rl}
\overline{\mathscr{P}}_\mathrm{err}&\geq1-(1-\mathscr{P}_\mathrm{out})(1-\epsilon),\label{eq05}\\
&\geq\mathscr{P}_\mathrm{out}+\epsilon-\epsilon\mathscr{P}_\mathrm{out}.\label{eq06}
\end{IEEEeqnarray}
We need to carefully manage the power $P_m$, so that $\mathscr{P}_\mathrm{out}$ can satisfy \eqref{eq06}.
In this case, the $m^\mathrm{th}$ sub-channel is successfully enabled to provide URLLC service, and the user assigned to this sub-channel is counted as a URLLC user.
It can be referred from \eqref{eq06} that to satisfy URLLC requirement, the outage probability requirement can be given by:
\begin{equation}\label{eq07}
\mathscr{P}_\mathrm{out}<\overline{\mathscr{P}}_\mathrm{out}=(\overline{\mathscr{P}}_\mathrm{err}-\epsilon)/(1-\epsilon),
\end{equation}
and $\mathscr{P}_\mathrm{out}$ would decrease with the increase of transmission power $P_m$.
To facilitate our study, we define the minimum power to enable a sub-channel as follows:
\begin{defn1}\label{def01}
The minimum power to enable URLLC service (denoted by $P_m^\perp$) is defined as the minimum power to achieve the outage probability requirement $\overline{\mathscr{P}}_\mathrm{out}$, and satisfies
\begin{equation}\label{eq08}
\mathscr{P}(R_m(P_m^\perp)<\overline{R})=\overline{\mathscr{P}}_\mathrm{out}.
\end{equation}
\end{defn1}
$P_m^\perp$ is dependent on the accuracy of CSIT, and it will be discussed later how to determine $P_m^\perp$. Based on $P_m^\perp$, the power allocation problem to maximize the number of URLLC users (denoted by $K$) can therefore be formulated as
\begin{IEEEeqnarray}{rl}\label{eq09}
\underset{P_0,P_1,...,P_{M-1}}{\max}~&K=\sum_{m=0}^{M-1}\mathbf{1}(P_m\geq P_m^\perp),\IEEEyessubnumber\label{eq09a}\\
\mathrm{s.t.}~&P_m\geq 0,\IEEEyessubnumber\label{eq09b}\\
&\sum_{m=0}^{M-1}P_m\leq MP,\IEEEyessubnumber\label{eq09c}
\end{IEEEeqnarray}
where $\mathbf{1}(\cdot)$ stands for the indicator function \cite{Digital_Comm}.

It is easy to know that this optimization problem is different from conventional throughput-oriented systems, where the transmitter carefully manages the transmission power based on its awareness of individual user PER. 
On the other hand, \eqref{eq09} is also different from current works for URLLC that focus on exploiting the diversity gain \cite{wang2021massivemimo,9120745}. Here the allocation should be based on the minimum power to enable a URLLC-user, which is related to the channel gain of the sub-channel. This motivates our following discussion in Sec. \ref{secIII}-\ref{secIV}.
To facilitate our measurement when $M$ is different, the user capacity is defined as
\begin{equation}\label{eq10}
\Gamma\triangleq(K)/(M).
\end{equation}
In Sec. \ref{secIV}, the results of $\Gamma$ will be further demonstrated.

\section{Power Allocation for URLLC-User Maximization}\label{secIII}
In this section, it is first proved that provided perfect CSIT, the user capacity can be maximized through a power-sorting method. Moreover, assuming imperfect CSIT, the proposed approach can still be used to optimize the user capacity for the pessimistic case, where the reliability of individual user is guaranteed.

\subsection{Power Allocation for URLLC-User Maximization Provided Perfect CSIT}\label{secIIIa}

When the transmitter is perfectly aware of the channel, the outage state of a sub-channel is certain to the transmitter (i.e., $\mathscr{P}_\mathrm{out}\in\{0,1\}$). Correspondingly, the outage probability requirement is set to $\overline{\mathscr{P}}_\mathrm{out}=0$.
In this case, the transmitter can perfectly calculate $R_m$ based on the knowledge of $h_m$ and $a_m$, and $P_m^\perp$ can be given by \textit{Lemma \ref{lem01}}.

\begin{lem}\label{lem01}
Given perfect CSIT, the value of $P_m^\perp$ can be calculated from the following equation:
\begin{equation}\label{eq11}
\overline{R}=R_m(P_m^\perp).
\end{equation}
\end{lem}  
\begin{proof}
Based on \textit{Definition \ref{def01}}, we have:
\begin{equation}\label{eq12}
\mathscr{P}(R_m(P_m^\perp)<\overline{R})=\overline{\mathscr{P}}_\mathrm{out}=0.
\end{equation}
From \eqref{eq02}, it is easy to know that $R_m(P_m)$ increases monotonically with $P_m$. Hence, the minimum power to satisfy \eqref{eq12} means
\begin{equation}\label{eq13}
\overline{R}=R_m(P_m^\perp).
\end{equation}	
With \eqref{eq12} and \eqref{eq13}, \textit{Lemma \ref{lem01}} is proved.
\end{proof}
Benefited from the monotonicity of $R_m(P_m)$, $P_m^\perp$ can be calculated using line search. This is a low-complexity algorithm and can be conducted in parallel for each sub-channel. So the calculation of $P_m^\perp$ does not bring significant calculation latency to the power allocation. If the power required for URLLC service is lower than the total power (i.e., $\sum_{m=0}^{M-1}P_m^\perp\leq MP$), the user capacity can be easily maximized by allocating $P_m^\perp$ to each of the sub-channel (i.e., $\Gamma=1$). However, such is usually not the case, particularly some of the sub-channels are in deep fading. In this case, we propose a power-sorting approach, where the sub-channels with lower $P_m^\perp$ are preserved to maximize the user capacity. Without loss of generality, it is assumed the sorted sequence of $P_m^\perp$ is given by:
\begin{equation}\label{eq14}
P_0^\perp\leq P_1^\perp\leq...\leq P_{M-1}^\perp.
\end{equation}
\begin{thm}\label{thm01}
Given the sorted power sequence \eqref{eq14}, the optimal solution to \eqref{eq09} is given by
\begin{IEEEeqnarray}{lr}\label{eq15}
P_m=\left\{
\begin{aligned}
&P_m^\perp,&0\leq m \leq K^\star-1,\\
&0,&K^\star\leq m\leq M-1,
\end{aligned}
\right.
\end{IEEEeqnarray}
\end{thm}
where $K^\star$ satisfies
\begin{equation}\label{eq16}
\sum_{m=0}^{K^\star-1}P_m^\perp\leq MP<	\sum_{m=0}^{K^\star}P_m^\perp.
\end{equation}
\begin{proof}
Assume the system can achieve a larger number of URLLC users (denoted by $K^\dagger$) than $K^\star$ (i.e., $K^\dagger-1\geq K^\star$). 	
Based on the sorted power-sequence \eqref{eq14}, the minimum power consumption when enabling $K^\dagger$ sub-channels for URLLC service is $\sum_{m=0}^{K^\dagger-1}P_m$. Therefore, we have the following inequality:
\begin{equation}\label{eq17}
\sum_{m=0}^{K^\dagger-1}P_m\geq\sum_{m=0}^{K^\star}P_m>MP.
\end{equation}
It is shown in \eqref{eq16} that for any $K^\dagger>K^\star$, the total power consumption will exceed the power budget $MP$.
This proves the solution in \eqref{eq14} and \eqref{eq15} is optimal to problem \eqref{eq09}.
\end{proof}
The value of $K^\star$ can be obtained by calculating $\sum_{m=0}^{K-1}P_m^\perp$ for $K=0,1,...,M-1$ and compare them to $MP$. This does not bring significant latency to the power allocation, either. 
In this case, the maximized user capacity is $\Gamma=(K^\star)/(M)$.
Moreover, it is worth mentioning that the enabled URLLC users are consuming minimum power to achieve individual reliability requirement. This means the power-sorting approach can also achieve the best transmit-power efficiency when maximizing the user capacity.

\subsection{Power Allocation for URLLC-User Maximization Provided Imperfect CSIT}
When the CSIT is imperfect, the transmitter is only aware of a corrupted version ($\hat{h}_m$) of the channel. In this case, $R_m(P_m)$ is a random variable to the transmitter, whose stochastic property is dependent on $\hat{h}_m$. 
In such a case, the transmitter needs to obtain $P_m^\perp$ based on $\hat{h}_m$ and $\overline{\mathscr{P}}_\mathrm{out}\in[0,1]$.
To facilitate our study, it is assumed that the CSIT uncertainty is white Gaussian for each sub-channel (e.g., \cite{1427697}):
\begin{equation}\label{eq18}
\hat{h}_m=h_m+e_m,~e_m\sim\mathcal{CN}(0,\sigma_\mathrm{e}^2)
\end{equation}
where $e_m$ stands for the CSIT uncertainty.
It is also assumed that the transmitter is aware of $\sigma_e^2$ to handle the CSIT uncertainty.
Then, $a_m=|\hat{h}_m-e_m|^2$ is a non-central $\chi^2$ distributed variable to the transmitter.
In the ideal case, the transmitter aims to find a threshold value $a_m^\perp$ such that 
\begin{equation}\label{eq19}
\mathscr{P}(a_m<a_m^\perp)=\overline{\mathscr{P}}_\mathrm{out}.
\end{equation}
Then, $P_m^\perp$ can be obtained from the following equation through line search (similar to \textit{Lemma \ref{lem01}}):
\begin{equation}\label{eq20}
\log_2\left( 1+\frac{a_m^\perp P_{m}^\perp}{N_0} \right)-\sqrt{\frac{V_{m}}{L}}\frac{\mathcal{Q}^{-1}(\epsilon)}{\ln 2}=\overline{R}.
\end{equation}
{Then, the outage requirement can be satisfied:
\begin{equation}
\mathscr{P}(R_m(P_m^\perp)<\overline{R})=\mathscr{P}(a_m<a_m^\perp)=\overline{\mathscr{P}}_\mathrm{out}.
\end{equation}
However, a low-complexity approach to computing $a_m^\perp$ is not available in the literature. Alternatively, we use the Chernoff lower bound (denoted by $a_{m,\text{Cher}}^\perp$) to obtain a pessimistic evaluation of $a_m^\perp$ with reasonable tightness \cite{wang2021massivemimo}.
\begin{lem}\label{lem02}
Given the imperfect CSIT $\hat{h}_m$, $\sigma_e^2$ and $\overline{\mathscr{P}}_\mathrm{out}$, the Chernoff bound $a_{m,\text{Cher}}^\perp$ can be obtained by line-searching $a_{m,\text{Cher}}^\perp\in(0,|\hat{h}_m|^2+\sigma_e^2)$ to solve the following equation 
\begin{equation}\label{eq21}
\overline{\mathscr{P}}_\mathrm{out}=f(a_{m,\text{Cher}}^\perp)=\frac{\exp(a_{m,\text{Cher}}^\perp t-\frac{-|\hat{h}_m|^2t}{1+\sigma_e^2t})}{1+\sigma_e^2t},
\end{equation}
where $t$ is an auxiliary variable given by
\begin{equation}\label{eq22}
t=\frac{\sigma_e^2+\sqrt{\sigma_e^4+4a_{m,\text{Cher}}^\perp|\hat{h}_m|^2}}{2\sigma_e^2a_{m,\text{Cher}}^\perp}-\frac{1}{\sigma_e^2}.
\end{equation}
\end{lem}
Due to the space limit, the proof is abbreviated here and please refer to Sec. III in \cite{wang2021massivemimo} for detailed proof. By substituting $a_{m,\text{Cher}}^\perp$ to \eqref{eq20}, $P_m^\perp$ can be obtained such that
\begin{equation}
\mathscr{P}(R_m(P_m^\perp)<\overline{R})=\mathscr{P}(a_m<a_{m,\text{Cher}}^\perp)\leq\overline{\mathscr{P}}_\mathrm{out}.
\end{equation}
Then, the rest of power allocation can be done by repeating the power-sorting approach introduced in \textit{Theorem \ref{thm01}}.
}

\section{Numerical Results and Discussion}\label{secIV}

\begin{figure}[t]
	\centering
	\includegraphics[scale=0.28]{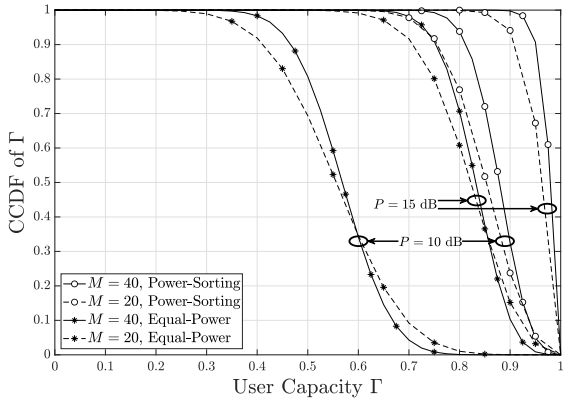}
	\caption{CCDF of the user capacity for the equal-power allocation and the proposed power allocation approach provided perfect CSIT regarding the number of sub-channels $M$ and the power per sub-channel $P$.}	\label{fig1}	
\end{figure} 

Numerical evaluations are carried out to analyze the performance of the proposed lightweight power-sorting approach in terms of user capacity as well as transmit-power efficiency. Specifically, the transmit-power efficiency is measured through average power consumption per URLLC user.
The channel is assumed to be Rayleigh fading in appreciation of its popularity in wireless communication research \cite{9120745,9107489}.
For the sake of simplicity, it is assumed $N_0=1$ and $P$ is normalized by $N_0$.
To deliver URLLC service, a $256$-bit packet needs to be transmitted in $0.5$-ms with a PER of $\overline{\mathscr{P}}_\mathrm{err}=10^{-5}$. This means the data rate requirement is $\overline{R}=(256~\mathrm{bit})/(0.5~\mathrm{ms})=512~\mathrm{kbps}$.
We consider a special case of FDMA: orthogonal frequency-division multiple-access (OFDMA) to be adopted for URLLC downlink \cite{3GPPTS38912}.
Each sub-channel is assigned one sub-carrier with the sub-carrier spacing to be $240$ kHz. Therefore, the symbol length $L=(0.5~\text{ms})\cdot(240~\text{kHz})=120$.
The sub-carrier spacing is chosen to be $240$ kHz so that the symbol length will be higher (compared to throughput-oriented systems, such as $15$ kHz in LTE systems \cite{3GPPTS36201}) to reduce the finite-length penalty on the achievable rate.
Three baselines are considered to in our numerical study.
\begin{itemize}
\item \textbf{Equal-Power Allocation:} the equal-power allocation stands for the case where the transmitter is not aware of any CSIT at all and treats each sub-channel equally ($P_m=P$).
\item \textbf{Waterfilling Allocation:} the waterfilling allocation stands for the case of throughput-oriented design where the system throughput is maximized ($P_m=\max(0,\lambda-\frac{N_0}{a_m})$, see \cite{Tse2005}).
\item \textbf{Equal-iSNR Allocation:} the equal-iSNR allocation stands for the case where the transmitter tries to enable each sub-channel by force ($P_m=\frac{MP}{\sum_{m=0}^{M-1}\frac{1}{a_m}}$).
It is presented to show the performance of enabling all sub-channels when the diversity gain is not available (in contrary to current works for URLLC).
\end{itemize}
The numerical study is divided into two examples:

\textit{Numerical Example 1:}
The objective of this example is to compare the proposed power-sorting approach to the baselines provided perfect CSIT in terms of user capacity as well as average power consumption per URLLC user.
In this case, $\epsilon=\overline{\mathscr{P}}_\mathrm{err}=10^{-5}$ and $\overline{\mathscr{P}}_\mathrm{out}=0$.
Fig.\ref{fig1} demonstrates the comparison of user capacity $\Gamma$ between the power-sorting approach and the equal-power allocation through their complementary cumulative distribution function (CCDF). It can be observed that the power-sorting approach can provide a significant improvement of around $0.25$ on $\Gamma$ when $P=10$ dB. 
As $P$ increases to $15$ dB, the improvement becomes around $0.12$. Such decrease is because as $P$ increases, each sub-channel is more likely to provide URLLC service with the average power. Nevertheless, the improvement is still impressive.  
Moreover, it can also be observed that the CCDF of the power-sorting approach is slightly steeper than the equal-power one. When $M$ increases from $20$ to $40$, this phenomenon becomes more significant. This is because compared to the equal-power allocation the power-sorting algorithm can compensate the power of sub-channels need less power than average to enable to those with poorer channel gain. Moreover, those sub-channels with very poor channel gain are abandoned, since activating those sub-channels cost significant higher power than others.

\begin{figure}[t]
	\centering
	\includegraphics[scale=0.28]{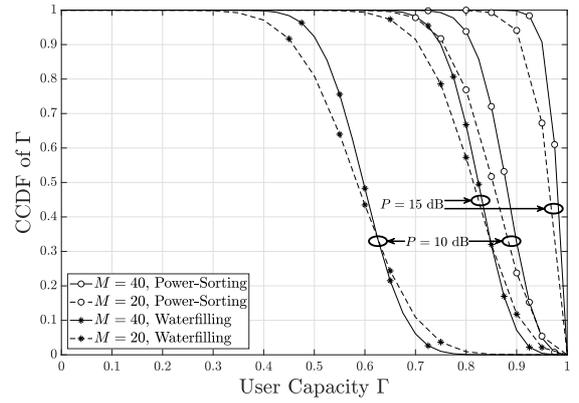}
	\caption{CCDF of the user capacity for the waterfilling allocation and the proposed power allocation approach provided perfect CSIT regarding the number of sub-channels $M$ and the power per sub-channel $P$.}	\label{fig2}	
\end{figure}

Fig. \ref{fig2} demonstrates the comparison of user capacity between the power-sorting approach and the waterfilling approach.
It can be observed that the power-sorting approach provides an improvement of around $0.18$ when $P=10$ dB and $0.15$ when $P=15$ dB.
It is not surprising that the waterfilling approach achieves even worse performance compared to the equal-power allocation when $P=15$ dB. This is because the idea of the waterfilling approach is to allocate more power to the sub-channels with good channel gain. When maximizing the number of URLLC users, such will polarize the iSNR across sub-channels and even decrease the user capacity. When $P=10$ dB, the waterfilling approach is slightly better than the equal-power allocation. This is because the waterfilling approach can abandon those sub-channels with poor channel gain and compensate the power to other sub-channels. Such principle actually coincides with the power-sorting approach, and therefore brings improvement on the waterfilling approach when $P$ is low.

\begin{figure}[t]
	\centering
	\includegraphics[scale=0.28]{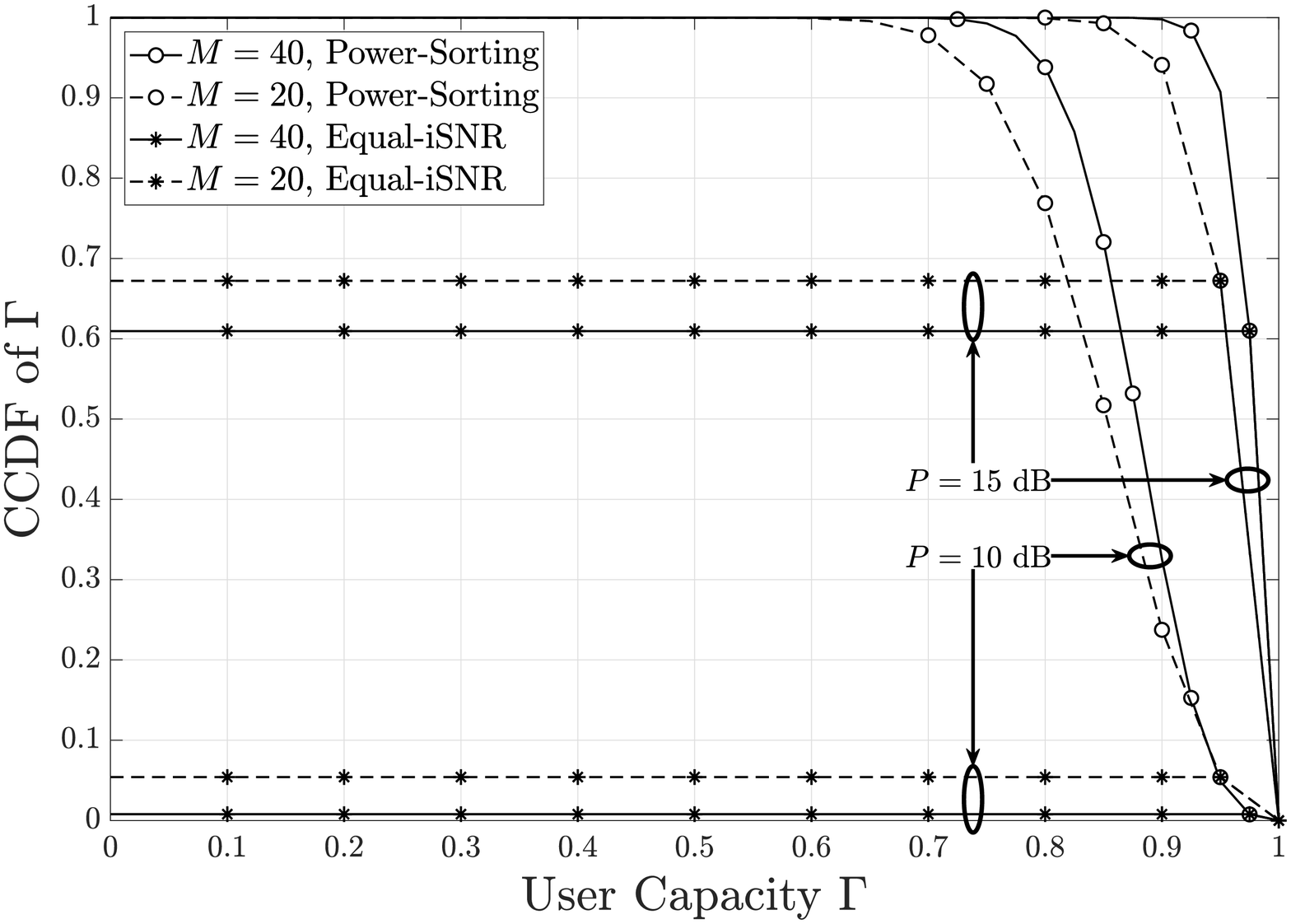}
	\caption{CCDF of the user capacity for the equal-iSNR allocation and the proposed power allocation approach provided perfect CSIT regarding the number of sub-channels $M$ and the power per sub-channel $P$.}	\label{fig3}	
\end{figure}

Fig. \ref{fig3} demonstrates the comparison of user capacity between the power-sorting approach and the equal-iSNR approach. The equal-iSNR approach actually has only two kinds of results: all sub-channels can successfully provide URLLC service, or all sub-channels fail to provide URLLC service. When $P$ increases from $10$ dB to $15$ dB, the success probability increases from less than $0.1$ to more than $0.6$. However, the power-sorting approach can achieve the same probability to enable all sub-channels. This is because the condition to enable all sub-channels is the same for both approaches (i.e., $\sum_{m=0}^{M-1}P_m^\perp\leq MP$). 
When $\Gamma<1$, the power-sorting approach significantly outperforms the equal-iSNR allocation. It can be observed in Fig. \ref{fig3} that $\Gamma\geq0.6$ for the power-sorting approach when $P=10$ dB, and $\Gamma\geq0.7$ when $P=15$ dB. This means the power-sorting approach can achieve a more stable user capacity than the equal-iSNR allocation. This again implies that when the power consumption to enable a sub-channel is too high, it is better to abandon this sub-channel. Different from the equal-power and waterfilling approach, the equal-iSNR achieves only worse performance when $M$ increases from $20$ to $40$. This is reasonable, since as $M$ increases, there is more likely to exists a sub-channel with very poor channel gain to degrade the system performance.

\begin{figure}[t]
	\centering
	\includegraphics[scale=0.28]{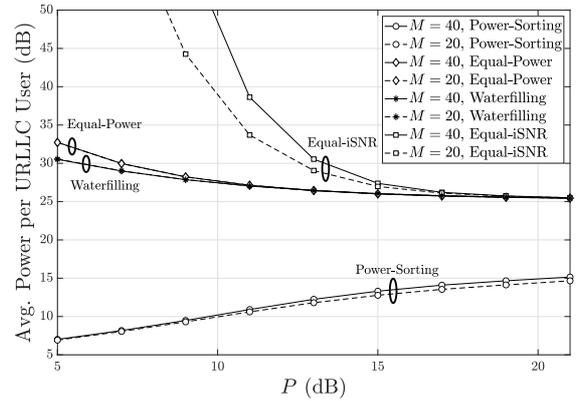}
	\caption{Comparison of average power consumption per URLLC user to the baselines provided perfect CSIT regarding the number of sub-channels $M$ and the power per sub-channel $P$.}	\label{fig4}	
\end{figure}

Fig. \ref{fig4} demonstrates the comparison of average power consumption per URLLC user between the power-sorting approach and the baselines. Apart from better user capacity, another reason for this improvement is that the power-sorting approach can save the power when the user capacity is already maximized. 
When $P$ decreases, such improvement can increase to more than $20$ dB. This is because the $P$ is low, the user capacity is more dependent on proper power allocation approach. While when $P$ is high, more sub-channels can enable themselves with average power, and the advantage of power allocation slightly decreases.

\textit{Numerical Example 2:}
The objective of this example is to investigate the performance when the CSIT is imperfect. The waterfilling and equal-iSNR allocation are not included in this example, as they do not manage the CSIT uncertainty. This means the outage probability due to overestimating the channel is $0.5$, which is too high for URLLC. The equal-power allocation has the same performance as before, and is therefore abbreviated here. 
To simplify the calculation, it is assumed that $1-\epsilon\approx1$ (since $\epsilon<\overline{\mathscr{P}}_\mathrm{err}\ll1$) \cite{wang2021massivemimo}. In this case, \eqref{eq07} can be approximated as:
\begin{equation}
\overline{\mathscr{P}}_\mathrm{out}\approx\overline{\mathscr{P}}_\mathrm{err}-\epsilon.
\end{equation}
Specifically, we consider the case where $\epsilon=\overline{\mathscr{P}}_\mathrm{out}=0.5\times10^{-5}$. 
Such symmetric setup is because setting a very small $\epsilon$ will lead to severe finite-length performance penalty in \eqref{eq02}; and setting a very small $\overline{\mathscr{P}}_\mathrm{out}$ makes the Chernoff bound $a_m^\perp$ in \textit{Lemma \ref{lem02}} too small, where both cases result in a high $P_m^\perp$. 

\begin{figure}[t]
	\centering
	\includegraphics[scale=0.28]{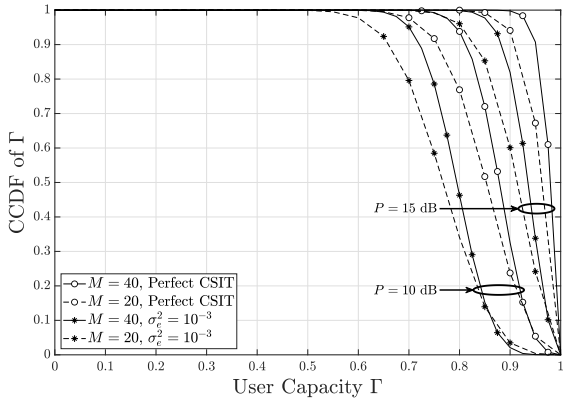}
	\caption{CCDF of the user capacity for the proposed power allocation approach provided perfect or imperfect CSIT regarding the number of sub-channels $M$ and the power per sub-channel $P$.}	\label{fig5}	
\end{figure}

 Fig. \ref{fig5} demonstrates the comparison of user capacity for the power-sorting approach between the case of perfect CSIT and $\sigma_e^2=10^{-3}$. It can be observed that the case of $\sigma_e^2=10^{-3}$ only has around $0.07$ degradation of user capacity when $P=10$ dB. When $P=15$ dB, such degradation reduces to around $0.04$. This means the proposed power-sorting approach can preserve remarkable performance at the presence of CSIT imperfection.  
 
\begin{figure}[t]
	\centering
	\includegraphics[scale=0.28]{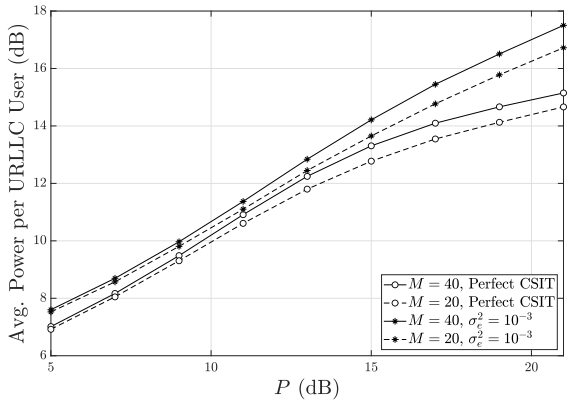}
	\caption{Comparison of average power consumption per URLLC user provided perfect or imperfect CSIT regarding the number of sub-channels $M$ and the power per sub-channel $P$.}	\label{fig6}	
\end{figure}

 Fig. \ref{fig6} demonstrates the comparison of average power consumption per URLLC user between the case of perfect CSIT and $\sigma_e^2=10^{-3}$. It can be observed that as $P$ increases from $5$ dB to $21$ dB, the increase of power consumption grows from around $0.5$ dB to around $2.5$ dB. This means the power-sorting approach can preserve reasonable average power consumption at the presence of imperfect CSIT. The reason that the performance gap becomes larger as $P$ increases is that when $P$ is high, more sub-channels with lower channel gain are enabled. But using Chernoff bound for the channel-gain will lead to more power consumption, particularly when $|\hat{h}_m|$ is already small.

\section{Conclusion}
In this paper, a novel power-sorting approach for power allocation has been proposed to maximize the user capacity for the FDMA-URLLC downlink with random channel assignment. Provided perfect CSIT, the proposed approach has proven to ensure maximized URLLC users. Provided imperfect CSIT, the proposed approach could still achieve remarkable user capacity by using Chernoff bound as a pessimistic evaluation of the channel gain. 
It has been demonstrated that the power-sorting approach can increase the user capacity for more than $0.12$ (out of $1$) compared to the equal-power, waterfilling and equal-iSNR allocation in Rayleigh fading channels. Moreover, the power-sorting approach has also been shown to achieve more than $10$ dB improvement on the average power consumption per URLLC user. 
Provided imperfect CSIT, the power-sorting approach can preserve remarkable user capacity (less than $0.07$ of degradation) with power consumption increase of less than $2.5$ dB.

\section*{Acknowledgement}
This work is funded by the 5G Innovation Centre and the 6G Innovation Centre.

\balance	

\ifCLASSOPTIONcaptionsoff
\newpage
\fi

\bibliographystyle{IEEEtran}
\bibliography{Bib_URLLC,Bib_Precoding,Bib_Else}		
\end{document}